\documentstyle[seceq,epsf,subfigure]{ptptex}

\makeatletter
\@ifundefined{lesssim}{\def\lesssim{\mathrel{\mathpalette\vereq<}}}{}
\@ifundefined{gtrsim}{\def\gtrsim{\mathrel{\mathpalette\vereq>}}}{}
\def\vereq#1#2{\lower3pt\vbox{\baselineskip1.5pt \lineskip1.5pt
\ialign{$\m@th#1\hfill##\hfil$\crcr#2\crcr\sim\crcr}}}
\makeatother

\title{
Is Dark Energy the Only Solution to the Apparent Acceleration
of the Present  Universe?
}

\author{
Hideo {\sc Iguchi}$^{1}$,
Takashi {\sc Nakamura}$^{2}$
and Ken-ichi {\sc Nakao}$^{3}$
}

\inst{
$^{1}$Department of Physics, Tokyo Institute of Technology, Tokyo 152-8550, Japan
\\
$^{2}$Yukawa Institute for Theoretical Physics, 
Kyoto University, Kyoto 606-8502, Japan\footnote{Present address: Department of Physics, Kyoto University, Kyoto 606-8502, Japan.}\\
$^{3}$Department of Physics, Osaka City University, Osaka 558-8585, Japan
}

\recdate{
August 2, 2002
}

\abst{
Even for the  observed luminosity distance $D_L(z)$, 
which suggests the existence of dark energy, 
we show that an inhomogeneous dust universe solution without dark energy is
possible in general. Future observation of $D_L(z)$ for
$ 1 \lesssim z < 1.7$  may confirm or refute this possibility.
}

\begin{document}

\maketitle
\section{Introduction}

Recent measurements of the luminosity distance $D_L(z)$ using
Type Ia supernovae
\cite{{Schmidt:1998ys},{Riess:1998cb},{Perlmutter:1999np}} 
suggest that an
accurate value of $D_L(z)$ may be obtained in the near future. 
In particular, SNAP 
\cite{snap} should provide us the luminosity distances of $\sim$2000 Type Ia 
supernovae with an accuracy of a few percent up to $z\sim1.7$ every year.  
Also, from observation of the first Doppler peak of
the anisotropy of the CMB,  it is suggested that the universe is
flat, \cite{ber00,lange00} and this may be proved  in the future from 
obervations by MAP and
Planck. Under the assumption of the homogeneity and isotropy of
our universe,  these observations suggest that
dark energy is dominant at present. 
In an attempt to determine the nature of 
dark energy, many arguments have been given. \cite{Weinberg:2000yb}
Recently, some mechanisms to account for the observed tiny but finite 
dark energy are proposed. \cite{Yokoyama:2001ez,Yokoyama:2002ts}
However, at present we do not
have a firm and reliable theoretical basis to investigate
 such a small
energy scale compared with the Planck scale. In short, 
the nature of dark energy under the assumption of the 
homogeneity and isotropy of our universe
is still a great mystery.

From the observed isotropy of the CMB, assuming that 
we are not in a special part
of our universe, the universe should be homogeneous. However,
if our position in the universe is
special, the universe might  be inhomogeneous, although
the CMB is isotropic.
Such  cosmological models have been constructed using  spherically symmetric
models in which we are   near the symmetric center.
Some authors have considered such models to interpret the SNIa data
 for small $z$, \cite{Celerier:2000hp}  as well as large $z$ 
assuming a void
structure \cite{{Tomita:2000a},{Tomita:2001a},{Tomita:2001b}} 
to avoid dark energy. 
Such   possibilities may be regarded as absurd.   
However, our point of view in this paper 
is to construct a possible inhomogeneous
dust universe derived from  the observed  $D_L(z)$. If such a model
 is consistent with present observational results,  
the inhomogeneous universe should be examined more
seriously, because the dark energy solution is also absurd in the
sense that it is $\sim 120$ orders of magnitude smaller than the
Planck scale. In short, we suppose that 
the question we need to answer to roughly reduces to the following:  
Which is more absurd, dark
energy or an inhomogeneous universe? In the former case, there is no
reliable theory to examine the problem at present, 
while  the latter case can be studied in the frame-work 
of known theories.  We would like to point out that it is not
taste but, rather, future observations that will confirm either dark energy or 
an inhomogeneous universe.

The analysis of high redshift supernovae gives us
the luminosity distance-redshift relation $D_{L}(z)$ along the
observational past null cone up to $z \sim 1$. 
\cite{{Schmidt:1998ys},{Riess:1998cb},{Perlmutter:1999np}}
The data fit well with   $D_{L}(z)$ in the homogeneous and
isotropic  universe with  $\Omega_m = 0.3$ and
$\Omega_\Lambda = 0.7$ given by
\begin{equation}
 \label{D_L}
 D_{L}(z) = \frac{1}{H_0}(1+z)
       \int_0^z \frac{dz'}{\sqrt{\Omega_m (1+z')^3 +\Omega_\Lambda}}.
\end{equation}
In this paper  we assume that  $D_{L}(z)$ is given by 
Eq. (\ref{D_L}) with  $\Omega_m = 0.3$ and
$\Omega_\Lambda = 0.7 $ for $z \lesssim 1$. This is done for the sake of
simplicity to make the arguments clearer. In particular, we do not wish to
claim that
 $D_{L}(z)$ with $\Omega_m = 0.3$ and $\Omega_\Lambda = 0.7 $ has been 
confirmed.
While  $D_{L}(z)$ for $ 1 \lesssim z < 1.7$ is not certain even at present
and will be obtained in the future, for example,  by SNAP.  
Since the scale factor $a$ obeys  
\begin{equation}
\frac{\ddot{a}}{a}=-\frac{4\pi}{3}(\rho + 3p),
\end{equation}
$D_{L}(z)$ with $\Omega_m = 0.3$ and $\Omega_\Lambda = 0.7$ 
implies that the present universe is accelerating,
while for the dust universe ($p=0$),  $a$ should be
decelerating. Therefore it may be concluded that
observations is inconsistent with the inhomogeneous
dust model. However, the point is that to determine  $D_{L}(z)$,
we are observing  Type Ia supernova events that occurred at past times in
spatial positions separated from us. In the inhomogeneous universe
model, the time dependence of  $a$ at a point separated from us 
differs from that of our position, 
so that we may obtain an apparent accelerating universe
even though the dust universe is decelerating locally.

Before ending this introduction, we comment on some other
works relevant to this paper. The inhomogeneous scenario is not 
the only alternative to dark energy.
Giving up the assumption that the cosmic substratum is composed of perfect 
fluids, bulk pressures that differ from the kinetic 
pressure can be allowed. The assumption
of an effective anti-friction force leads to a model that has only one 
dark component (CDM) and is consistent with the CMB and SNIa data.
\cite{Zimdahl:2001zm}
Also, there is an approach somewhat related to that presented in this paper
(though with different motivation) that has been used.
\cite{Avelino:2001a,Avelino:2001b,Avelino:2001c}

\section{Formulation}

The line element of a spherically
symmetric dust universe is given by
\begin{equation}
 ds^2 = -dt^2 + \frac{(R'(t,r))^2}{1+2E(r)r^2}dr^2 + R^2(t,r) d\Omega^2,
\end{equation}
 where the prime indicates differentiation with respect to $r$. 
The solution to the Einstein equations  is known as the 
Lema\^{\i}tre-Tolman-Bondi
(LTB) spacetime, \cite{Lemaitre:1933,Tolman:1934,Bondi:1947} given by 

\begin{eqnarray}
 \dot{R} &=& \sqrt{\frac{2M(r)}{R}+2E(r)r^2}, \label{Rdot} \\
 4\pi\rho (t,r) &=& \frac{ M'}{R^2R'} \label{rho},
\end{eqnarray}
 where the dot indicates differentiation with respect to $t$. 
The solution of Eq.\ (\ref{Rdot}) is given by 
\begin{equation}
 \label{R(t,r)}
 R(t,r) = \frac{M}{\epsilon (r)r^2}\phi (t,r), ~~~ 
 t-t_B(r)=\xi(t,r)\frac{M}{(\epsilon (r)r^2)^{3/2}},
\end{equation}
where 
\begin{equation}
 \epsilon(r)r^2 = \left\{ \begin{array}{ccc} 
                           2 E(r)r^2, & (E(r)>0)\\
                            1, & (E(r)=0)\\ 
                           -2E(r)r^2,  & (E(r)<0)
                           \end{array}\right. 
\end{equation}
and
\begin{equation}
 \phi = \left\{ \begin{array}{c} 
                           \cosh\eta-1 , \\
                            \frac{\eta^2}{2} ,\\ 
                           1-\cos\eta ,
                           \end{array}\right. ~~
\xi = \left\{ \begin{array}{ccc} 
                           \sinh\eta-\eta , & (E(r)>0)\\
                            \frac{\eta^3}{6} , & (E(r)=0)\\ 
                           \eta-\sin\eta . & (E(r)<0)
                           \end{array}\right. 
\end{equation}
In the general solutions of the LTB models,  there are three 
arbitrary functions 
$M(r), E(r)$ and $t_B(r)$.  $M(r)$ is regarded as the gravitational mass
function, and we can set $M(r)=M_0r^3$, redefining  $r$. $t_B(r)$ 
corresponds to the local
BigBang time. $E(r)$ determines the local curvature radius 
or the local specific energy.
The functions $t_B(r)$ and  $E(r)$ should be chosen 
to reproduce the observed  $D_L(z)$. 
This means that we have only one constraint for two arbitrary functions.

The observational past null cone is specified in the form, 
$t={\hat t}(r)$. We denote the areal radius $R$ on $t={\hat t}(r)$ 
by $\cal R$. Then, by Eq. (\ref{Rdot}), we can regard  
$\dot R$ on $t={\hat t}(r)$ as a function of $\cal R$, $E$ and $r$:
\begin{equation}
   \dot{R}({\hat t}(r),r)=
   {\cal R}_{1}({\cal R},E,r)\equiv \sqrt{{2M_0r^3\over {\cal R}}
   +2Er^{2}}. \label{calR1}
\end{equation}
By differentiating Eqs. (\ref{Rdot}) and (\ref{R(t,r)}),  
$R'$ and ${\dot R}'$ on $t={\hat t}(r)$ can be expressed 
as functions of $\cal R$, $\hat t$, $E$, $E'$, $t_{B}$, $t_{B}'$ 
and $r$:\footnote{If $E(r)=0$, we should omit the terms proportional
to $E'$ in Eqs. (\ref{calR2}) and (\ref{calR3}).}

\begin{eqnarray}
   R'({\hat t}(r),r)
&=&{\cal R}_{2}\left({\cal R},{\hat t},E,E',t_{B},
   t_{B}',r\right) \nonumber \\
&\equiv& 
   -\left( {\cal R} -\frac{3}{2}\left[{\hat t}-
   t_B\right]{\cal R}_{1}  \right) \frac{E '}{E} 
-{\cal R}_{1} t_B' \nonumber \\
 &&+ \frac{\cal R}{r}, \label{calR2}
\end{eqnarray}
and
\begin{eqnarray}
   \dot{R'}(\hat{t}(r),r) &=&
   {\cal R}_{3}\left({\cal R},{\hat t},E,E',t_{B},
   t_{B}',r\right) \nonumber \\
&\equiv&\frac{1}{2} \left({\cal R}_{1} - 3
   \frac{M_0 r^3}{{\cal R}^2} \left[t-t_B\right]\right)
   \frac{E'}{E}  \nonumber \\
&+& \frac{M_0 r^3}{{\cal R}^2} t_B'+\frac{{\cal R}_{1}}{r}. \label{calR3}
\end{eqnarray}

The observational past null cone  $t=\hat{t}(r)$ satisfies 
\begin{equation} 
 \label{dhattdr}
  \frac{d\hat{t}}{dr} = -\frac{{\cal R}_{2}
({\cal R}, {\hat t},E,E',t_{B},t_{B}',r)}{\sqrt{1+2Er^2}}.
\end{equation}
The redshift $z(r)$ along the past null cone is given by
\begin{equation}
 \label{dzdr}
 \frac{dz}{dr}=\frac{1+z}{\sqrt{1+2Er^2}}
 {\cal R}_{3}({\cal R},{\hat t},E,E',t_{B},t_{B}',r) .
\end{equation}
The total derivative of $\cal R$ on the past null cone is written
\begin{eqnarray}
 \label{dRdr}
 \frac{d{\cal R}}{dr}
 =\left(1-\frac{{\cal R}_{1}({\cal R},E,r)}{\sqrt{1+2Er^2}}\right)
  {\cal R}_{2}({\cal R},{\hat t},E,E',t_{B},t_{B}',r).
\end{eqnarray}
Our basic equations are Eqs. (\ref{dhattdr})$-$(\ref{dRdr}). 
These three equations can be regarded as a system of 
first-order ordinary differential equations for three of 
the five functions  ${\cal R}(r)$, $\hat t(r)$, $E(r)$, $t_B(r)$ and $z(r)$. 
In order to integrate these equations, we must
specify two conditions on these five functions. 
The luminosity distance $D_L(z)$ is related to ${\cal R}$
\cite{Partovi:1984} as
\begin{equation}
 {\cal R}=\frac{D_L(z)}{(1+z)^2}.
\end{equation}
As mentioned above, we assume that $D_{L}(z)$ is given by 
Eq. (\ref{D_L}). We will specify one further
condition on $E$, $t_B$ or a combination of them. 

\section{Results}
\subsection{Results of the BigBang time inhomogeneity}

We first consider a pure BigBang time inhomogeneity. In this case,
the curvature function $E(r)$ is set to a constant value.
From Eqs. (\ref{dzdr}) and (\ref{dRdr}), we have equations for $z(r)$
and the  Big-Bang time function $t_B(r)$. 
The model is specified by $\Omega_0\equiv 2M_0/H_0^2$, which 
is the present central density, $3M_0/4\pi$, divided by 
the present central critical density,
$\rho_{\mbox{\scriptsize crit}}={3H_0^2}/{8\pi}$, 
where $H_0$ is the present central Hubble parameter, and 
we set it to unity.   
We numerically integrated these two differential equations from $r=0$ 
for ten $\Omega_0$  from 0.1 to 1.0. 
The initial conditions are given by $z=0$ and $t_{B}=0$. 

\begin{figure}
  \begin{center}
    \leavevmode
    \epsfysize=2.2in\epsfbox{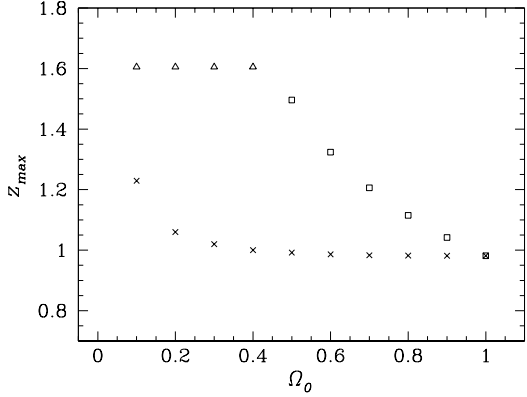}    
  \caption{ Plots of the maximum redshift of the time that  either of the 
  inequalities in  Eq. (\ref{conditions})
is violated as a function of the present density parameter. The open
  triangles and the open squares correspond to the BigBang time
  inhomogeneity. The crosses correspond to the curvature inhomogeneity.
   }
 \label{fig:zmax}
  \end{center}
 \end{figure}

From Eq. (\ref{rho}),  $R'>0$ for positive density, while  
from Eq. (\ref{dzdr}),
$~\dot{R'}>0$ for monotonically increasing $z(r)$,  so that
the integration is terminated when either of the inequalities
\begin{equation}
 \label{conditions}
 R'>0 ~~~\mbox{or} ~~~\dot{R'}>0
\end{equation}
is violated.
 In Fig. \ref{fig:zmax}, we display the relation between the parameter 
$\Omega_0$ and the redshift at the time that the integration is terminated.
For small value of $\Omega_0$, $\Omega_0=$ 0.1 -- 0.4 (open triangles),
 shell-crossing
singularities appear when $d{\cal R}/dz=0$. For large value of $\Omega_0$, 
$\Omega_0=$ 0.5 -- 
1.0 (open square), the second condition in Eq. (\ref{conditions}) is violated
first. This  occurs when ${\cal R}=2M$.

 \begin{figure}
  \begin{center}
    \leavevmode
    \epsfysize=2.2in\epsfbox{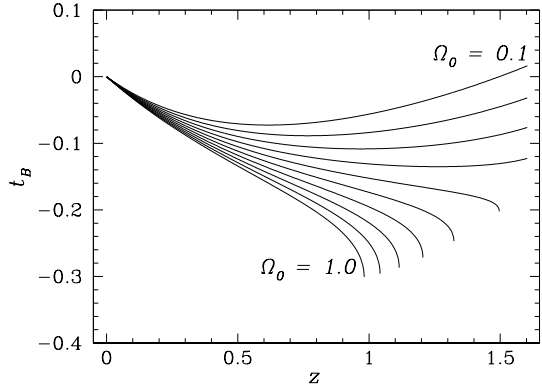}    
  \caption{ Plots of the BigBang time functions as functions of the
redshift $z$ .
   }
 \label{fig:tB}
  \end{center}
 \end{figure}

Figure \ref{fig:tB} plots the BigBang time functions $t_B$ for each
$\Omega_0$. 
For all $\Omega_0$, the  BigBang time functions $t_{B}$ 
decrease as $z$ increases up to $z \sim 0.5$.  
This result is related to the fact that 
the expansion of our universe appears to be accelerating 
up to $z\sim0.5$. In inhomogeneous models, 
an apparent acceleration is realized  
if the recession velocity of mass shells does not increase rapidly
along the observational past null cone as in the case of a
homogeneous and isotropic universe filled with dust.
To construct such a situation in our model, we need to prepare an
older shell, i.e., one that is more decelerated by gravity, for more distant
shells on the past null cone. This is the reason that the function $t_B$
decreases. 

\begin{figure}
  \begin{center}
    \leavevmode
    \epsfysize=2.2in\epsfbox{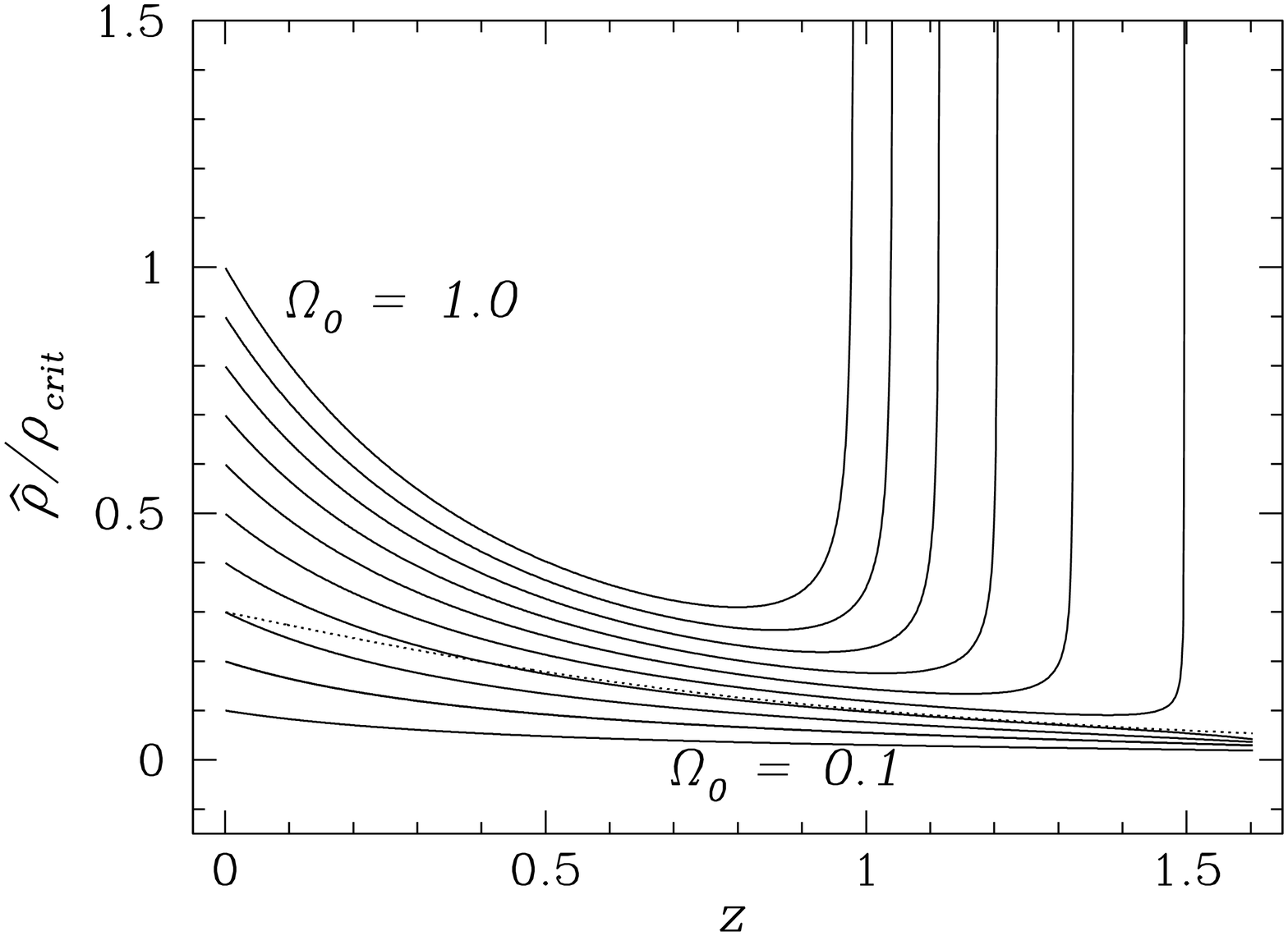}    
  \caption{ Plots of the redshift space density $\hat{\rho}$ 
   divided by the central critical density. The dotted curve represents 
   the $\Omega_m=0.3, ~ \Omega_{\Lambda}=0.7$ homogeneous model. 
   }
 \label{fig:tB_rho}
  \end{center}
 \end{figure}

In Fig. \ref{fig:tB_rho} we plot 
the redshift space density,
\begin{equation}
 \hat{\rho}(z) = \rho \frac{4\pi R^2 R' dr}{4\pi z^2 dz}
               = \Omega_0 \frac{r^2}{z^2}\frac{dr}{dz} \rho_{\mbox{\scriptsize crit}},
\end{equation}
along the past null cone. Observations of the mass distribution
along the past null cone would give us this density profile.

\subsection{Results of curvature inhomogeneity}

Next, we consider the pure curvature inhomogeneity. In this case the
BigBang time function $t_B(r)$ is set to zero. From
Eqs. (\ref{dhattdr}) -- (\ref{dRdr}) we obtain three
differential equations for the three variables $z(r)$, $E(r)$ and
$\hat{t}(r)$.
We numerically integrated these three differential equations from
$r=0$.  The initial conditions are given by $z=0$, $E=(1-\Omega_{0})/2$ 
and 
\begin{equation}
 \hat{t}(0)=\frac{\Omega_0}{2}\frac{(\sinh \eta_0  -\eta_0)}
{(1-\Omega_0)^\frac{3}{2}},
\end{equation}
where
\begin{equation}
 \eta_0 = \ln \left( \frac{2-\Omega_0}{\Omega_0} +
 \sqrt{\left(\frac{2-\Omega_0}{\Omega_0}\right)^2 -1}\right). 
\end{equation}
The present central cosmological time $\hat{t}(0)$ and $\eta_0$ are
obtained from Eq. (\ref{R(t,r)}). 

In Fig. \ref{fig:zmax}, we show the relation between the parameter 
$\Omega_0$ and the redshift at the time that the integration is 
terminated (cross
marks). For the case of a curvature
inhomogeneity, it is found that the second condition in Eq.\
(\ref{conditions}) is violated first. \cite{Kurki-Suonio:1992}

\begin{figure}
  \begin{center}
    \leavevmode
    \epsfysize=2.2in\epsfbox{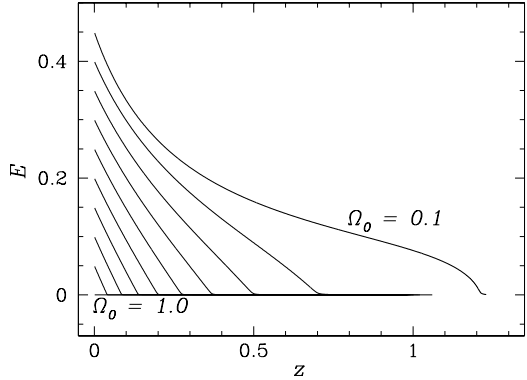}    
  \caption{ Plots of the curvature functions.
   }
 \label{fig:E}
  \end{center}
 \end{figure}

Figure \ref{fig:E} displays the curvature functions $E$ for 
variations values of
$\Omega_0$. We can see $E$ decreases as $z$ increases, except in the 
$\Omega_0=1.0$ case. 

The result that $E$ decreases is consistent with  the
apparent acceleration.  The value of $E$ determines the
specific energy of the dust elements, so that the ``initial'' velocity
is smaller for more distant shells. This causes
apparent acceleration, because the velocity at $r=0$ can be largest.

\begin{figure}
  \begin{center}
    \leavevmode
    \epsfysize=2.2in\epsfbox{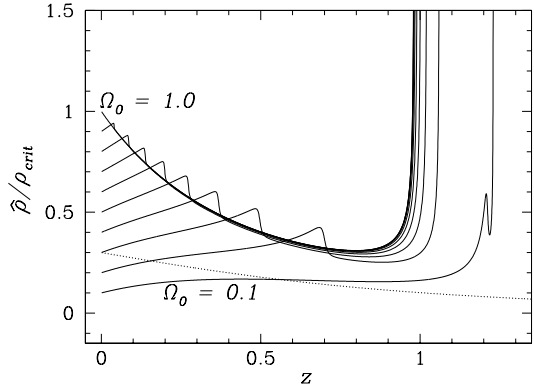}    
  \caption{ Plots of the redshift space density divided by the present central 
  critical density. The dotted curve represents
   the $\Omega_m=0.3, ~ \Omega_{\Lambda}=0.7$ homogeneous model. 
   }
 \label{fig:curv_rho}
  \end{center}
 \end{figure}

Figure \ref{fig:curv_rho} 
displays the redshift space density $\hat{\rho}$ along 
the past null cone as a function of $z$. 

\section{Summary and discussion}

In this paper we have constructed inhomogeneous dust models
without dark energy. We find that these models are consistent with
the observed $D_L(z)$ up to $z=1$, as from Fig. 1 no
difficulties are encountered 
up to $z\sim 1$ for any set of parameter values in both the BigBang
time inhomogeneity and curvature inhomogeneity cases.
For $z > 1$, we have difficulties in our inhomogeneous dust
models. 
Recently, the SNIa at a redshift of 
$\sim 1.7$  was found  \cite{{Gilliland:1999ne},{Riess:2001gk}}
with rather large uncertainties. 
However, only  a single SNIa at a redshift of 
$\sim 1.7$ is not enough to construct an accurate  $D_L(z)$, 
although  that result  seems to rule out the `grey-dust' hypothesis. 
In addition, the results of a 
recent investigation of the effect of gravitational lensing
on this SNIa suggests that the grey-dust model may be consistent 
with the observational data. \cite{Mortsell:2001,Gunnarsson:2001}
If future observations confirm $D_L(z)$ up to $z\sim 2$ 
with $\Omega_m \sim 0.3$ and $\Omega_\Lambda \sim 0.7$ , 
it can be concluded that
our inhomogeneous dust models are incompatible with the observations
and that some form of dark energy is likely to exist.
However, if future observations confirm  that  $D_L(z) $ for $z > 1$ 
is not consistent with Eq. (\ref{D_L}), the plausibility of 
our inhomogeneous dust models should be studied more extensively.
 In such a case,  the first Doppler peak as well as the higher ones
will give us another constraints on the inhomogeneous universe models.

It may believed that the existing observations for $0<z<1$, such as 
 (i) evolution of cluster abundance, (ii) lensing rate, and
(iii) ages of stellar populations, already rule out the inhomogeneous
models. 

Using the cluster temperature evolution data 
for  $0.3< z < 0.8$, it was found that the best-fit value 
for $\Omega_m$ is $\Omega_m=0.45 \pm 0.1$ for open 
universe and  $\Omega_m=0.3 \pm 0.1$ for flat universe. \cite{Donahue:1999} 
However, recent analysis shows that the systematic error is comparable to
the statistical error. \cite{Voit:2000}
Therefore, we may say that  $0.1 < \Omega_m <0.5$ for  $0.3< z < 0.8$ data.
It is not clear whether the Press-Schechter formalism
can be applied to our inhomogeneous  models. 
One possible estimate of cluster abundance 
could be based on the locally homogeneous
approximation. As we know, massive cluster evolution is very 
sensitive to matter density. It seems that a model whose local 
density parameter $\Omega_m$ differs greatly from the best-fit value
would not be able to explain the observed cluster
evolution. The pure curvature inhomogeneity case with 
$\Omega_0 \gtrsim 0.2$ may not survive, because it is 
approximated by a flat universe at large $z$. Also the BigBang time
inhomogeneity case with $\Omega_0 \sim 1.0$ cannot survive. 
However, it can be expected that the pure BigBang time inhomogeneity with
$\Omega_0 \sim 0.5$ and the pure curvature inhomogeneity with
$\Omega_0 \sim 0.1$ will predict the observed cluster abundances. 

The estimate of the lensing rate and the distribution of the separation 
of the images depend on the model used for the 
mass distribution of the lensing object and the luminosity function of the 
source objects as well as the cosmological parameters. 
However, it has been shown that the dependence on the lens 
model and parameters is much stronger 
than that  on the cosmological parameters. 
\cite{Takahashi:2001,Chiba:2001} In addition, the mass distribution of the 
lensing objects should depend strongly on the baryon density $\Omega_b$.
\cite{Kochanek:2001} Therefore, we 
conclude that the estimate of the cosmological parameters from the 
lensing rate and the distribution of the separation of the
images is difficult at present, and for this reason,  we cannot rule out the 
inhomogeneous model.

\begin{figure}
  \begin{center}
    \leavevmode
    \epsfysize=2.2in\epsfbox{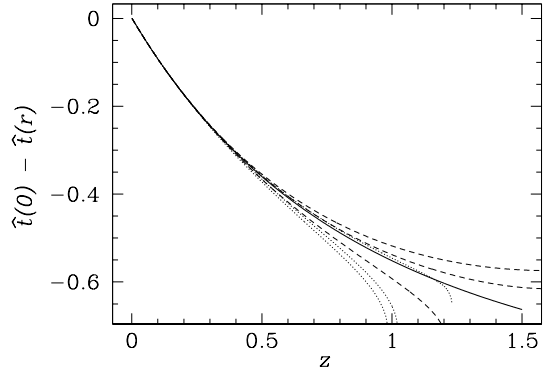}    
  \caption{ Plots of the look back time along the past null cone.
  The solid curve represents the homogeneous $\Omega_m = 0.3$,
  $\Omega_\Lambda = 0.7$ case. The broken and dotted lines denote 
  the pure BigBang time and the pure curvature inhomogeneity cases 
  with $\Omega_0=0.1, 0.3, 0.7$ in descending order, respectively.
   }
 \label{fig:times}
  \end{center}
 \end{figure}

As shown in Fig. \ref{fig:times}, the look back times along the past 
null cone differ little 
between the inhomogeneous model and the corresponding homogeneous   
model with cosmological constant for  $z < 0.5$. For  $z \sim 1$, 
a difference appears,
but some of the inhomogeneous models do not differ greatly from the 
homogeneous model even in that case. The ages of a 
stellar population could not be used to 
distinguish  the inhomogeneous model from  the homogeneous one.

We have found that the model dependence, including
 various undetermined parameters, and the observational uncertainty
 are much larger than the dependence on the cosmological parameters.
 Therefore we believe that these observations cannot easily rule out the 
inhomogeneous model.

Before finishing, we give a brief comment on how we can be positioned away 
from the center of the symmetry.
A displacement from the center would correspond to a dipole mode of 
CMB. Therefore we can be positioned $\sim 50$ Mpc away from the center.

In conclusion, dark energy is not the  only possible solution of the 
apparent acceleration
of the present  universe, as inhomogeneous dust models can also
account for current observations.

\section*{Acknowledgements}

We are grateful to K. T. Inoue for helpful discussions.
This work was partially supported by  Grants-in-Aid for Scientific
Research (No. 11217 and  Nos.11640274 and 09NP0801) from the 
Japanese Ministry of Education, Culture, Sports, Science and Technology.

\end{document}